\documentclass[aps, prl, amsmath, reprint, showkeys, superscriptaddress]{revtex4-2}
\usepackage{wrapfig}
\usepackage{float}
\usepackage{graphicx, tikz}
\usepackage[caption=false]{subfig}
\usepackage{hyperref}
\usepackage{xcolor}
\usepackage{algorithm}
\usepackage{algpseudocode}

\newcommand{\norm}[1]{\left\lVert#1\right\rVert}
\newcommand{\abs}[1]{\left\lvert#1\right\rvert}
\newcommand{\x}{\boldsymbol{x}}
\newcommand{\C}{\boldsymbol{C}}
\renewcommand{\O}{\mathcal{O}}

\newcommand{\sub}[2]{#1_{\mathrm{#2}}}

\begin{document}

\title{Adaptive Node Positioning in Biological Transport Networks}

\author{Albert Alonso}
\altaffiliation{Authors contributed equally.}
\affiliation{Niels Bohr Institute, University of Copenhagen, Denmark}
\author{Lars Erik J. Skjegstad}
\altaffiliation{Authors contributed equally.}
\affiliation{Niels Bohr Institute, University of Copenhagen, Denmark}
\author{Julius B. Kirkegaard}
\email[Correspondence email address:]{juki@di.ku.dk}
\affiliation{Niels Bohr Institute, University of Copenhagen, Denmark}
\affiliation{Department of Computer Science, University of Copenhagen, Denmark}

\date{\today}

\begin{abstract}
Biological transport networks are highly optimized structures that ensure power-efficient distribution of fluids across various domains, including animal vasculature and plant venation.
Theoretically, these networks can be described as space-embedded graphs, and rich structures that align well with observations emerge from optimizing their hydrodynamic energy dissipation.
Studies on these models typically use regular grids and focus solely on edge width optimization.
Here, we present a generalization of the hydrodynamic graph model which permits additional optimization of node positioning.
We achieve this by defining sink regions, accounting for the energy dissipation of delivery within these areas, and optimizing by means of differentiable physics.
In the context of leaf venation patterns, our method results in organic networks that adapt to irregularities of boundaries and node misalignment, as well as overall improved efficiency. 
We study the dependency of the emergent network structures on the capillary delivery conductivity and identify a phase transition in which the network collapses below a critical threshold.
Our findings provide insights into the early formation of biological systems and the efficient construction of transport networks.
\end{abstract}
\maketitle

Transport networks are ubiquitous in nature and in living systems.
The efficiency of such networks is crucial for the evolutionary fitness of organisms such as those observed in leaf venation~\cite{mcculloh_water_2003, ronellenfitsch_global_2016, katifori2012quantifying} and blood vasculature systems~\cite{murray1926physiological, katifori2012quantifying, kirkegaard_optimal_2020}, and emerges in complex systems such as river networks~\cite{konkol2022interplay} and human transport systems~\cite{lonardi2021designing, lonardi2023bilevel, leite_similarity_2024}.
Understanding the structure and morphogenesis of such network structures has been facilitated by studying energetically optimal solutions to static, hydrodynamic networks~\cite{bohn_structure_2007, katifori_damage_2010, corson_fluctuations_2010, hu_adaptation_2013}.

The optimization of hydrodynamic transport networks is traditionally approached as an edge-optimization problem~\cite{bohn_structure_2007, kirkegaard_optimal_2020, ronellenfitsch_global_2016, hu_adaptation_2013}, assuming systems where network nodes serve as sinks for their local area.
Thus, optimizing the energy dissipation leads to optimal edge conductivities~\cite{bohn_structure_2007}, while the positions of the nodes themselves are considered fixed.
In principle, this approach can be used to model any bounded system if enough nodes are used.
However, for a finite number of nodes confined within a bounded system, it is evident that the node placement itself influences the optimality of the fluid delivery system.
In many systems, the finite number of nodes is a physical fact and must be imposed, e.g., due to a lower bound on the vein thicknesses determined by the capillary size.

\begin{figure}[b]
    \centering
    \includegraphics[width=8cm]{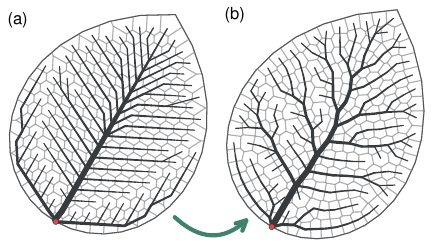}
    \caption{
    Optimization and relaxation on an inclined leaf ($\varphi_0=60^\circ$, $\beta=0.4$, $N=175$).
    Non-suppressed edges and delivery regions of each sink node are shown.
    The source is marked with a red dot.
    (a)~Result after transport network optimization solving Eq.~\eqref{eq:c-ode-growth} on a hexagonal grid.
    (b)~Result after relaxation, where the network optimizes $P=\sub{P}{t} + \sub{P}{d}$.
    }
    \label{fig:showcase}
\end{figure}

Here, we consider bounded systems of finite nodes and generalize the hydrodynamic model to have well-defined optima both in edge conductivities and node positioning.
Specifically, we model a bounded leaf venation network and let the boundary represent the leaf margin.
Each node represents a source or sink (we set a single node at the leaf base as the source, and let all other nodes be sinks) and each edge represents a vein between two nodes.
Fig.~\ref{fig:showcase}a shows the optimal solution obtained for a fixed, hexagonal grid in a bounded system.
The imposition of the domain boundary and the inability of the nodes to adapt positionally results in non-uniform areas associated with each node.
As fluid dissipation will typically be proportional to area, this thus implies that the sinks become non-uniform.
By removing the spatial constraint on the node positions, we find more energy-efficient solutions, which yield networks that appear organic and more consistent with networks observed in nature (Fig.~\ref{fig:showcase}b).

The power of the transport network is given by~\cite{bohn_structure_2007}
\begin{equation}\label{eq:transport_power}
    \sub{P}{t} = \sum_{e \in \text{edges}} \left( \frac{F_e^2}{C_e} + c_t \, C_e^\gamma \right) L_e,
\end{equation}
where $F_e$ is the flux associated with edge $e$, which has length $L_e$ and conductivity $C_e$.
The constant $c_t$ defines the metabolic cost of maintaining the edge, which is equivalent to considering a system with a finite amount of resources.
$\gamma$ is a parameter that determines how the material cost scales with the conductivities, which we set to $\gamma=1/2$ in this Letter.
The flow is pressure-driven, such that the flux over a (directed) edge $e = i \rightarrow j$ is given by $F_e = (p_i - p_j) C_e/L_e$.
The pressures are indirectly determined by the need to satisfy Kirchhoff's law $\sum_{e \in i} \pm_e F_e = s_i$, where $s_i$ is the source/sink at node $i$ and the sign is set by the direction of the edge.
This can be solved efficiently with hardware acceleration~\cite{skjegstad_modeling_2024}.
Typically, the sink magnitudes are assumed to be equal.
However, once the nodes are permitted to move, these values must change as well.
Thus, taking the source $s_0 = 1$, we define $s_i = -A_i/\sum{A_i}$ ($i\geq1)$, where $A_i$ is the area surrounding a node, implicitly defined by the associated Voronoi cell.

In order for a transport system to have stable and well-defined optima in node positioning, the above model must be expanded to account for the power dissipation within the Voronoi cells.
To achieve this, we add a simple power-delivery term, $\sub{P}{d}$, to the conventional formulation for transport power~\cite{bohn_structure_2007}, and thus assume that the total power can be described by the sum of two contributions $P = \sub{P}{t} + \sub{P}{d}$.
With the goal of defining a self-consistent model that equally considers both transport and delivery costs, we take
\begin{equation}\label{eq:delivery_power}
    \sub{P}{d} = \sum_{i \in \text{sinks}}\frac{s_i^2}{\hat{C}_i} \langle \ell \rangle_i.
\end{equation}
This is analogous to the first term in the transport formulation (Eq.~\eqref{eq:transport_power}), but considers the power dissipation due to the delivery of the sink fluid $s_i$ over an average Voronoi distance $\langle \ell \rangle_i$.
$\hat{C}_i$ is the delivery conductivity, which effectively models a capillary system.
Physically, this term favors equally sized and isotropic sink areas.
We assume that the delivery conductivity is an intrinsic property of the material, such that $\hat{C}_i = \hat{C}$ is fixed and equal for all nodes.

\vspace{1em}
\paragraph{Optimization ---}
Our scheme to minimize $P$ consists of the combination of an edge conductivity optimization and a node relaxation process.
We employ a fully connected graph and initialize all edge conductivities to the same value.
For a given node positioning $\{ \boldsymbol{x}_i \}$, we optimize the conductivities by using an adaptation model~\cite{hu_adaptation_2013, ronellenfitsch_global_2016}
\begin{equation}\label{eq:c-ode-growth}
    \dfrac{\mathrm{d} C_e}{\mathrm{d} t} = \left( \dfrac{F^2_e}{C_e^{\gamma+1}} - \gamma c_t \right) C_e + c_0 e^{-\lambda t},
\end{equation}
where the term $F^2_e \mathbin{/} C_e^{\gamma+1}$ is the squared wall shear stress, and $\gamma c_t$ represents the optimal squared shear stress.
Furthermore, we include a growth term characterized by the area growth rate $\lambda$, which has proven to be a robust strategy to achieve better optima~\cite{ronellenfitsch_global_2016}. 
To facilitate smooth convergence, we also decrease exponentially the magnitude of the growth term $c_0(\tau)\propto e^{-\nu \tau}$ during the node positioning \textsl{relaxation process}.
Here, we use $\tau$ to denote the timescale of node relaxation and $t$ for the conductivity optimization.

Crucially, even though our transport network is fully connected, optimization results in a sparse, planar graph~(Fig.~\ref{fig:showcase}).
While networks with static nodes can simplify the incidence matrix by, for instance, employing Delaunay triangulation to account for sparse connections, we rely on the derivative information of the entire graph during the optimization to accurately determine the correct gradients for node displacement and to allow for the spontaneous creation and suppression of connections.

\begin{figure}[!tb]
    \centering
    \includegraphics[width=8cm]{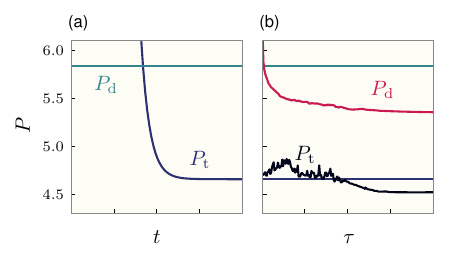}
    \vspace{-1.5em}
    \caption{
    Transport and delivery power terms (Eqs. \eqref{eq:transport_power} and \eqref{eq:delivery_power}) during optimization.
    (a)~Optimization of $\sub{P}{t}$ by using Eq~\eqref{eq:c-ode-growth}.
    (b)~Node relaxation process. At each timestep one complete optimization of $\sub{P}{t}$ is performed, resulting in an improvement of both power terms.
    }
    \label{fig:optimization}
\end{figure}

To optimize over $\{ \boldsymbol{x}_i \}$, we use gradient descent computed by automatic differentiation techniques~\cite{rall_arithmetic_1986, jax2018github}.
To propagate the gradients across the solution of the adaptation model in Eq.~\eqref{eq:c-ode-growth}, we employ a custom backpropagation method based on the implicit function theorem~\cite{bell_algorithmic_2008}, detailed in~\cite{supp}.
Importantly, the fact that we use Voronoi cells to define the sink magnitudes $s_i$ and the delivery distances $\langle \ell \rangle_i$ means that all terms vary continuously with node positions, yet it requires the ability to differentiate through Voronoi calculations.
We achieve this with a custom differentiable implementation of Voronoi tessellation~\cite{shumilin_method_2024, numerow_differentiable_2024, feng_cellular_2023}, clipped by the domain boundary \footnote{Code implementation is available at \url{https://github.com/kirkegaardlab/gradnodes}} (see~\cite{supp} for details).
Crucially, this procedure is parallelizable and amenable to hardware acceleration.
As Voronoi cells are guaranteed to be convex, we can split the integrals over these cells into triangles $T$, e.g.,
\begin{equation}\label{eq:average-distance-in-voronoi-cell}
    \langle \ell \rangle_i = \dfrac{1}{A_i} \sum_{T \in \text{vor}_i}
        \int\!\!\!\int_{T} \norm{\boldsymbol{x}_i - \boldsymbol{x}} \, \mathrm{d}A,
\end{equation}
over which the integrals can be evaluated analytically (see~\cite{supp}).
We note that our optimization schemes are local optimizers and will generally not find global optima.

\vspace{1em}

During the optimization of conductivities, delivery cost remains fixed~(Fig.~\ref{fig:optimization}a).
On the other hand, during node placement relaxation, we see that while delivery costs are monotonically decreasing, transport costs experience a phase of stochasticity where the efficiency is reduced~(Fig.~\ref{fig:optimization}b).
Nevertheless, the total power continuously decreases, and after some iterations both the transport power $\sub{P}{t}$ and the delivery power $\sub{P}{d}$ reach lower values than those achievable in the regular grid, showing that the terms are not opposing criteria and that the simultaneous optimization yields networks that show increased power efficiency in both.

\begin{figure}[!tbh]
    \centering
    \includegraphics[width=8cm]{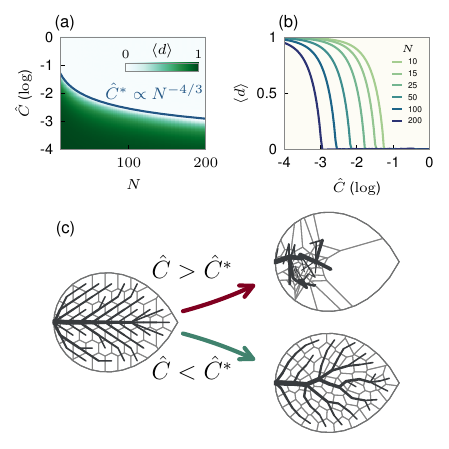}
    \caption{
    Network collapse as a function of effective delivery conductivity.
    (a)~Phase diagram of (normalized) average nearest-neighbor distance for different $\hat{C}$ and network sizes in the one-dimensional system.
    The line shows the scaling of $\hat{C}$ at constant $\Omega$, (analytically obtained from the scaling analysis).
    (b)~Individual network sizes from (a) showing the sudden but continuous collapse.
    (c)~Resulting networks with $\hat{C}$ above and below the critical value in two dimensions.
    }
    \label{fig:phase-transition}
\end{figure}

\vspace{1em}
\paragraph{Network stability ---}
The delivery term $\sub{P}{d}$ depends on the capillary conductivity $\hat{C}$.
To understand its influence on the network dynamics, we consider a one-dimensional system, where the nodes are solely connected to their adjacent nodes.
This removes the need for Voronoi calculations, where instead the average delivery distance is
\begin{equation}
    \label{eq:average-distance-1d}
    \langle \ell \rangle_i =  \frac{1}{2(x_R-x_L)}\left[ (x_R-x_i)^2 + (x_i-x_L)^2 \right],
\end{equation}
where $x_L$ and $x_R$ are the midpoints between the left and right neighbor, respectively~(see~\cite{supp}).
Figs.~\ref{fig:phase-transition}a--b show the mean node separation $\langle d \rangle$ in this 1D system confined to a fixed domain for increasing values of $\hat{C}$.
We observe a smooth phase transition:
When $\hat{C}$ is small, the energy is largely dominated by the delivery energy, which is minimal when the nodes are uniformly spread.
As we increase the delivery conductivity of the system, we observe a continuous change in the network until the network collapses~(Fig.~\ref{fig:phase-transition}b).

To understand the origin of this transition, we consider the contributions of each power term.
If the energy needed to transfer fluid from a sink node to the surrounding region is significantly greater than the energy required for transportation (small $\hat{C}$), the cost incurred by reducing the connections between nodes -- and thereby minimizing transport energy -- is too high to justify compromising the size discrepancies of the regions.
In contrast, when the delivery conductivity is sufficiently high to allow effectively energy-free fluid transfer from the sink to the region, the optimal solution is achieved by reducing the transport cost, i.e., moving all the nodes to the source, as a single node can then deliver fluid to the entire domain cheaply.
This leads us to study the transition using the dimensionless power ratio $\Omega = \sub{P}{t} / \sub{P}{d}$.

Scaling analysis of the one-dimensional system reveals that $\mathcal{O}(\Omega) = N^{{4}\mathbin{/}{3}}$~(see~\cite{supp}).
Since $\Omega \propto \hat{C}$, we observe that the abrupt collapse of the system in Fig.~\ref{fig:phase-transition}a aligns with the outcomes of the scaling analysis.
This analysis can be extended to the two-dimensional formulation, resulting in  $\mathcal{O}(\Omega) = N^{{5}\mathbin{/}{3}}$, which predicts that \textit{valid} relaxed networks only emerge below the critical $\hat{C}^\ast$, as shown in Fig.~\ref{fig:phase-transition}c.
Although more complex in two dimensions, the phase transition still occurs.

We note that the above scaling laws are derived for a fixed domain space.
If instead, we consider a 2D domain that increases with network size, e.g., $\mathcal{O}(L) = \sqrt{N}$ (i.e., a growing leaf), we find the resulting scaling to be $\mathcal{O}(\Omega) = N^{-{1}\mathbin{/}{3}}$.
Thus, for fixed domains, the more densely packed the network is, the lower the capillary conductivity must be in order for the formulation to hold.
If, however, the domain expands, the system can maintain the same effective conductivity during growth -- as expected of an intrinsic property.
For systems with a scaling domain, delivery costs become negligible at large network sizes, where instead the transport costs dominate completely.

\begin{figure}[!tb]
    \centering
    \includegraphics[width=8cm]{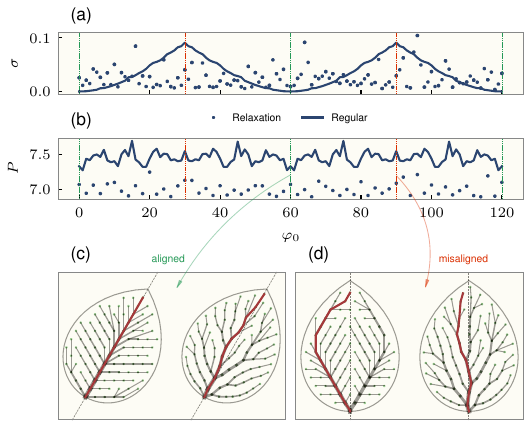}
    \caption{Model adaptability on differently aligned leaves.
    (a-b)~The deviation $\sigma$ (a), as well as power $P$ (b) are $60^\circ$-periodic for the regular grid, and non-periodic for the relaxation model.
    (c-d)~Resulting leaves for maximum alignment (c) and misalignment (d) are fundamentally different for the regular grid (c-d, left), but similar for the relaxation model (c-d, right).}
    \label{fig:adaptability-to-alignment}
\end{figure}

\vspace{1em}
\paragraph{Adaptability ---}
One constraint of regular lattice graphs is that their optimal solutions are dependent on the alignment of the main axis of the leaf with the grid.
Typically, this alignment is chosen so that the main axis follows the shortest path between neighboring nodes, resulting in network shapes where the leaf main branch is perfectly parallel with the main axis (Fig.~\ref{fig:showcase}a).
Another issue is the aforementioned non-uniformity of the sink areas along the leaf boundary.
Thus, for fixed grids, careful construction and alignment of the network boundary is required.
To illustrate this, we construct a leaf shape that can be rotated relative to the grid by using a simplified version of Gielis' superellipse equation~\cite{gielis_generic_2003}, as it describes a broad range of leaf shapes accurately~\cite{shi_general_2018}:
\begin{equation}
    r(\varphi; \varphi_0, \beta) = \Bigl( \abs{\cos\left(\tfrac{\varphi-\varphi_0}{4}\right)} + \abs{\sin{\left(\tfrac{\varphi-\varphi_0}{4}\right)}} \Bigr)^{-1{/}\beta}.
\end{equation}
Here $\varphi_0$ is the inclination of the major axis of the leaf, and $\beta$ is a shape parameter.

We quantify the discrepancies by rotating the leaf and measuring the power $P$ and the average deviation of the main branch with the grid, $\sigma = 1 \, - \, \bigl \langle c_i \, \mathrm{cos}( \theta_i \, - \, \varphi_0) \bigr \rangle / \bigl \langle  c_i \bigr \rangle$. Here, $\theta_i$ and $c_i$ are the angle and conductivity of the $i$'th edge of the main branch, respectively.
The main branch is found by greedily following the edges with the highest conductivity, from the source to a leaf node.
On the hexagonal grid, angles $\varphi_0 \equiv 0 \, (\text{mod} \,60^{\circ})$ correspond to maximum alignment of the leaf with the grid, whereas $\varphi_0 \equiv 30^{\circ} \, (\text{mod} \,60^{\circ})$ correspond to maximum misalignment.

Fig.~\ref{fig:adaptability-to-alignment}a shows the average main branch deviations for the outputs of the regular grid and the relaxation model.
For the regular grid, we see that the deviation is completely dependent on angle, with a minimum of $\sigma = 0$ at the maximum values of leaf alignment, as well as peaks at maximum misalignment.
The slightly non-smooth changes in deviation are caused by the discrete differences in the initializations of the node positions due to boundary clipping.
For the deviations in the relaxation model, there is no apparent dependency on angle, which indicates that our model is independent of initial boundary alignment, with the noisiness explained by the fact that the relaxation model output is the result of local optimization.
While zero deviation is never reached for the relaxation model (due to the stochastic nature of the optimization), the average deviation $\sub{\sigma}{avg} = 0.026$ is nonetheless lower than $\sub{\sigma}{avg} = 0.033$ for the regular grid.

The resulting power from the same data can be seen in Fig.~\ref{fig:adaptability-to-alignment}b.
The mean power is significantly lower for the relaxation model, which indicates that the solutions in general are more optimal.
We note that at maximum alignment we still see a lower power in the relaxation model, even though the regular grid output has lower deviation, which can be attributed to the fact that power is not solely determined by the structure of the main branch, but by the network in its entirety.

Figs.~\ref{fig:adaptability-to-alignment}c-d show illustrations of the output of the two models for maximum alignment and misalignment, respectively.
The regular-grid model yields two extremes where the main branch either completely follows the main axis (Fig.~\ref{fig:adaptability-to-alignment}c, left), or splits into two similarly sized branches (Fig.~\ref{fig:adaptability-to-alignment}d, right).
In contrast, all outputs from the relaxation model preserve the central main branch (Figs.~\ref{fig:adaptability-to-alignment}c-d, right).

\begin{figure}[!b]
    \centering
    \includegraphics[width=8cm]{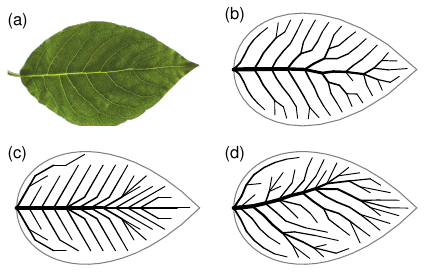}
    \caption{
    Comparison between a real leaf and model venation patterns ($\beta=0.25, N=100$).
    (a)~Sample of \textit{L. xylosteum} \cite{skjegstad_modeling_2024}.
    (b-d)~Model output from the (b)~relaxation model, (c)~regular grid, and (d)~sequential optimization.}
    \label{fig:real-leaves}
\end{figure}

\vspace{1em}
\paragraph{Vein curvature ---}
A characteristic feature of our resulting networks is the emergence of smoothly curving veins.
This is particularly noticeable in the branches extending from the main branch and leads to venation patterns that appear more organic compared to those produced by a regular grid.
Previous work relied on complementary initialization techniques in order to effectively mimic biological stochasticity, for example by using disordered tessellation grids~\cite{ronellenfitsch_global_2016}.
Such a method works by enforcing a repulsive potential between sink nodes that leads to evenly spaced positions in the domain.
This is approximately equivalent to independently optimizing $\sub{P}{d}$ in our formulation.

In Fig.~\ref{fig:real-leaves} we compare the resulting morphologies of a leaf that mimics the domain of \textit{L. xylosteum}~(Fig.~\ref{fig:real-leaves}a), and observe how venation patterns agree with biological observations in our formulation~(Fig.~\ref{fig:real-leaves}b), and a regular grid~(Fig.~\ref{fig:real-leaves}c), respectively.
In Fig.~\ref{fig:real-leaves}d we show the result of sequential optimization, i.e., one optimization of $\sub{P}{d}$ without accounting for transport cost, and one subsequent optimization of $\sub{P}{t}$.
We observe sub-optimal solutions where the venation patterns do not match those of the real leaf.
Instead, we find that the optimization of $\sub{P}{t}$ identifies approximately straight lines in the optimal node positions, and uses these to form the main branch.

Our observations indicate that incorporating coupled energy costs in the power formulation leads to solutions influenced by domain boundaries, closely resembling those found \textit{in vivo}~(Fig.~\ref{fig:real-leaves}b).
This is reproduced to a lesser extent when performing sequential optimization~(Fig.~\ref{fig:real-leaves}d) and is completely absent when relying on regular grids~(Fig.~\ref{fig:real-leaves}c).

\vspace{1em}

In this Letter, we have addressed the limitations of traditional transport network optimization models by allowing the optimization over the node positions themselves.
We have shown this to be a well-defined problem when incorporating the cost of resource delivery into the energy formulation, with resulting configurations that show adaptability to domain boundaries and misalignment.
The study has been enabled by exploiting a fully differentiable process including the Voronoi tessellation of the domain and the steady state of the conductivity adaptation model.
Our results demonstrate the advantages of node localization both internally and in adapting to external domain boundaries and allow the emergence of natural organic networks.
Our approach could find applications in the efficient design of human-engineered networks~\cite{barber_optimal_2008}.
We identified a phase transition as a function of the conductivity $\hat{C}$ of the delivery system, showing network collapse above a critical value, reminiscent of the well-known phase transition in $\gamma$ for static transport networks~\cite{bohn_structure_2007}.
Scaling analysis further reveals that physical networks remain stable during network growth only if the domain expands along with the network.
While our goal has been a minimal extension of the hydrodynamic network model, we note that a main insight is the addition of \textit{some} delivery power term and not necessarily precisely the one of Eq. \eqref{eq:delivery_power}.
For instance, similar phase transition behavior emerges from considering, e.g., $\langle \ell^2 \rangle$ (diffusion-limited costs).
Our approach is limited by the local behavior of gradient descent and the presence of many local optima.
Resulting patterns are thus sensitive to initial conditions even though the energy dissipation is similar between them.
This problem similarly prevails in edge optimization alone~\cite{bohn_structure_2007}.
Our findings pose an interesting question for further research:
how can appropriate \textit{local} feedback models~\cite{hu_adaptation_2013, ronellenfitsch_global_2016} be formulated that optimize $\sub{P}{t} + \sub{P}{d}$?
Finally, we note that in this Letter we do not consider fluctuations in the sink magnitudes or similar phenomena that can result in reticulate networks~\cite{katifori_damage_2010,corson_fluctuations_2010,waszkiewicz2024goldilocks}, the relevance of which becomes evident for larger network sizes.
This can be incorporated by choosing a suitable parameterization that allows for area-weighted sinks \cite{skjegstad_modeling_2024}.

\begin{acknowledgements}
This work has received funding from the Novo Nordisk Foundation, Grant Agreement No. NNF20OC0062047.
\end{acknowledgements}

\bibliography{references}

\onecolumngrid
\appendix
\newpage

\section{Spatially continuous sinks: integrals}

In this section, we provide a detailed analysis of the area computation of the region where the sink nodes deliver to and explore the integration of distance metrics in a triangle within the polygon.
Take a \textit{star convex} polygon (such as those that result from Voronoi tesselations),
and consider a triangle $ABC$ in that polygon:

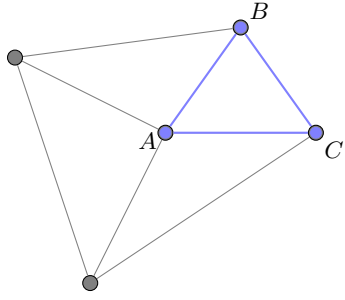
\begin{figure}[!h]
\centering
\begin{tikzpicture}
\foreach \point [count=\i from 0] in {(-1,0), (1,0), (-2,-2), (-3,1), (0,1.4)}
    \node[draw, circle, inner sep=2pt, fill=gray] (N\i) at \point {};
\foreach \i in {1, 2, 3} {
    \pgfmathtruncatemacro{\next}{\i+1}
    \draw[gray] (N0) -- (N\i) -- (N\next);
}
\draw[blue!50, fill] (N0) circle (2pt) node [black, left, yshift=-3] {$A$};
\draw[fill, blue!50] (N4) circle (2pt) node [black, above right] {$B$};
\draw[fill, blue!50] (N1) circle (2pt) node [black, below right] {$C$};
\draw[thick, blue!50] (N0) -- (N4) -- (N1) -- (N0);
\end{tikzpicture}
\caption{Diagram of a triangle of a Voronoi cell. $A$ is the sink node and $B$ and $C$ are vertices resulting from the tessaration.}
\end{figure}

The area of such a triangle is trivial to calculate, and thus the total area is:
\begin{equation}
    A = \sum_{T \in \text{triangles}} A_T,
\end{equation}
where $A_T$ is trivially computed with
\begin{equation}
    A_T = \frac{1}{2} \abs{(\boldsymbol{x}_{A} - \boldsymbol{x}_{B}) \times (\boldsymbol{x}_{A} - \boldsymbol{x}_{C})}.
\end{equation}

Likewise, we need the average distance from $A$ to all points in the polygon.
This becomes:
\begin{equation}
    \langle L \rangle = \frac{1}{A} \sum_{T \in \text{triangles}} \int_{A_T} || \boldsymbol{x}_A - \boldsymbol{x} || \, \mathrm{d}A.
\end{equation}
This integral is not easy, but can be done by using polar coordinates. We write
\begin{equation}
    \x(r, \theta) = \x_A + r \begin{pmatrix}
        \cos \theta \\ \sin \theta,
    \end{pmatrix}
\end{equation}
and the integral becomes
\begin{equation}
    \mathcal{I} = \int_{A_T} || \boldsymbol{x}_A - \boldsymbol{x} || \, \mathrm{d}A = \int_{\theta_1}^{\theta_2} \mathrm{d}\theta  \int_0^{R(\theta)} \mathrm{d} r \, r^2 = \frac{1}{3} \int_{\theta_1}^{\theta_2}  R(\theta)^3 \, \mathrm{d}\theta,
\end{equation}
where $R(\theta)$ is the distance from $A$ to the line $BC$ with angle $\theta$,
\begin{equation}
    R(\theta) = \frac{x_a(y_b - y_c) + x_b (y_c - y_a) + x_c (y_a - y_b)}{(y_c - y_b) \cos \theta + (x_b - x_c) \sin \theta},
\end{equation}
thus, our integral becomes
\begin{equation} \nonumber
    \mathcal{I} = \frac{1}{3} \left[ x_a(y_b - y_c) + x_b (y_c - y_a) + x_c (y_a - y_b) \right]^3 \int_{\theta_1}^{\theta_2}  \frac{1}{(\delta y \cos \theta - \delta x \sin \theta)^3} \, \mathrm{d}\theta,
\end{equation}
where $\delta x = x_c - x_b$ and $\delta y = y_c - y_b$.
This integral can be evaluated using  Weierstrass substitution to give
\begin{align} \nonumber
    \int \frac{1}{(\delta y \cos \theta - \delta x \sin \theta)^3} \, \mathrm{d}\theta =  &\frac{1}{\delta^3} \tanh^{-1} \! \left( \frac{\delta x + \delta_y \tan(\theta/2)}{\delta} \right) \\
    &+ \frac{\delta x \cos \theta + \delta y \sin \theta}{2 \delta^2 ( \delta y \cos \theta - \delta x \sin \theta)^2},
\end{align}
where we defined $\delta = \sqrt{\delta x^2 + \delta y^2}$. 
Note that
\begin{align}
    \tan(\theta/2) = \frac{\sin \theta}{1 + \cos \theta},
\end{align}
so this can all be written in terms of $\cos \theta$ and $\sin \theta$ (i.e. no need to actually calculate $\theta$).
Thus, the formula is complete by specifying
\begin{align}
    \cos \theta_1 = \frac{x_B - x_A}{|| \x_B - \x_A ||}, && \sin \theta_1 = \frac{y_B - y_A}{|| \x_B - \x_A ||}, \\
    \cos \theta_2 = \frac{x_C - x_A}{|| \x_C - \x_A ||}, && \sin \theta_2 = \frac{y_C - y_A}{|| \x_C - \x_A ||}.
\end{align}

As Weierstrass substitution has problems at $\theta = \pi$, the integral is only valid when the triangle is aligned with node $A$ west of nodes $B, C$ --- but this is always achievable by a simple rotation.

\newpage
\section{Solution for the 1-Dimensional case}

Here, we show the derivations and explanations to find the optimal distribution of $\x$ in the simplified case of a one-dimensional (1D) leaf.
We begin by considering a system with a single source $s_0 = 1$ at $x_0 = 0$, and $n$ delivery nodes positioned at locations $\x = \{ x_1, x_2, \dots, x_n \}$, which get constrained to $x\in [0, L]$.

Similar to the 2D case, the sink values $s_i$ represent the rate at which flow is absorbed or removed at each node $x_i$.
The sink value is calculated based on the distances between the neighbouring nodes, normalized by the system length $L$.
An exception to this is the first and last nodes, where the distance to the boundary is used instead.
\begin{equation}
s_i = - \frac{1}{2L} (x_{i+1} - x_{i-1}),
\end{equation}
with special attention to correct at $x_0$ and $x_n$.

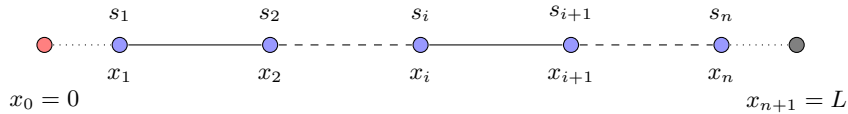
\begin{figure}[htb]
    \centering
\begin{tikzpicture}
    \draw[fill=red!50] (1,0) circle (0.1) node (init) {};
    \node[below] at (1,-0.5) {$x_0 = 0$};
    \foreach \x/\label in {2/1, 4/2, 6/i, 8/i+1, 10/n} {
        \draw[fill=blue!40, fill opacity=1] (\x, 0) circle (0.1) node (\label) {};
        \node[below] at (\x, -0.2) {$x_{\label}$};
        \node[above] at (\x, +0.2) {$s_{\label}$};
    }
    \draw (1) -- (2);
    \draw[dashed] (2) -- (i);
    \draw (i) -- (i+1);
    \draw[dashed] (i+1) -- (n);
    \draw[fill=gray] (11,0) circle (0.1) node (end) {};
    \node[below] at (11, -0.5) {$x_{n+1} = L$};
    \draw[dotted] (n) -- (end);
    \draw[dotted] (init) -- (1);
\end{tikzpicture}
    \caption{Diagram of the 1D transport networks with moving nodes (blue) and a fixed source (red).}
    \label{fig:enter-label}
\end{figure}

Next, we compute the flow values $F_i$ of the passing flow through each node.
Since we are in 1D, we assume that the nodes are only connected to the nodes on their sides.
Hence, the flow expression for each edge simplifies to
\begin{equation}
    F_{i+1} = F_{i}  - s_{i} = F_1 - \sum_{j\le i} s_j,
\end{equation}
given that $F_1 = 1$.

The conductivity at each node denoted $C_i$, is a function of the flow $F_i$.
We can find the optimal conductivity for this system by minimizing the transport power $P_t$ given by 
\begin{equation}
    P_{t} = \sum_{i \in \text{transport}} L_i \left( \frac{F_i^2}{C_i} + c_t C_i^\gamma \right),
\end{equation}
by setting $\frac{\partial P}{\partial C} = 0$, we find the conductivity at node $x_i$ to be:
\begin{equation}\label{eq:conductivity-optimal}
C_i = \left( \frac{F_i^2}{c_t \gamma} \right)^{\frac{1}{1+\gamma}}.
\end{equation}
This formula captures the non-linear dependence of conductivity on the square of the flow, modulated by the parameters $c_t$ and $\gamma$.

The total power $P$ of the system consists of two components: the transport power $P_t$ and the delivery power $P_d$.
The transport power is given by:
\begin{equation}
P_{\text{t}} = \sum_{i \in \text{transport}} L_i \left( \frac{F_i^2}{C_i} + c_t C_i^\gamma \right),
\end{equation}
where $L_i$ represents the length of the transport segment corresponding to node $x_i$, and the two terms inside the summation represent the power due to flow and the power related to the conductivity, respectively.

The delivery power, on the other hand, is the power used to deliver flow from the sink nodes to their surrounding areas, and it depends on the sink values $s_i$ and the average delivery distance $\langle \ell_i \rangle$.
The delivery power is expressed as:
\begin{equation}
P_{\text{delivery}} = \sum_{i \in \text{delivery}} \langle \ell_i \rangle \frac{s_i^2}{\hat{C}},
\end{equation}
where $\hat{C}$ represents the delivery conductivity, and $\langle \ell_i \rangle$ is the average distance over which the flow must be delivered.

The average delivery distance $\langle \ell_i \rangle$ for a node at position $x_i$ is computed using the positions of the left and right domain edges, denoted $x_L$ and $x_R$, respectively. The formula for the average delivery distance is the weighted average:
\begin{equation}
\langle \ell_i \rangle = \frac{(x_R-x_i)}{x_R-x_L}\cdot \dfrac{(x_R-x_i)}{2}+ \frac{(xi-x_L)}{(x_R-x_L)}\cdot \frac{(x_i - x_L)}{2} =  \frac{1}{2(x_R-x_L)}\left[ (x_R-x_i)^2 + (x_i-x_L)^2 \right]
\end{equation}
This formula accounts for the spatial distribution of the flow and the relative positions of the nodes within the system.

\newpage
\section{Scaling for different network sizes}
\subsection{One-dimensional network}
\label{subsec:one-dim-scaling}

We analyze the scaling of the total power $P$ given by
\begin{equation}
P = P_t + P_d,
\end{equation}
with respect to the number of elements $N$ on the 1 dimensional case using the expressions above.
For this, we assume the domain to be fixed with $N$.

Starting by the transport term, we have
\begin{equation}
P_\text{t} = \sum_{i \in \text{transport}} L_i\left(\frac{F^2}{C_i} + c_t C_i^\gamma \right),
\end{equation}
where $\O(F_i^2) \approx 1/N^2$, and thus $\O(C_i) = (\frac{1}{N^2})^{2/3}$ according to Eq.~\eqref{eq:conductivity-optimal}.
We also assume $c_t$ constant and $\gamma=0.5$ as in the main text.
Taking into account $L_i \propto \frac{L}{N}$, and splitting the terms we find:
\begin{equation}
\O(P_t) =\O(\Sigma)\ \O(L)\ \O(F^2)\ \O(C^{-1})  +  \O(\Sigma)\ \O(L)\ \O(C^{1/2}) = N^{-\frac{2}{3}} =  N^{-\frac{2}{3}} + N^{-\frac{2}{3}}.
\end{equation}

Hence, the transport network scales as $\O(P_t) = N^{-2/3}$.

In the case of the delivery term, given by
\begin{equation}
P_d = \sum_{i \in \text{delivery}} \langle \ell_i \rangle \frac{s_i^2}{\hat{C}},
\end{equation}
where $\langle \ell_i \rangle \propto \frac{L}{N}$, $s_i^2 \propto \frac{1}{N^2}$ and $\hat{C}$ is constant, giving:

\begin{equation}
\O(P_d) = \O(\Sigma)\ \O(\langle \ell \rangle)\ \O(s^2) =  N^{-2}.
\end{equation}

The ratio between the transport and cost terms to the delivery term scales as:
\begin{equation}
\O(\Omega) = \frac{\O(P_d)}{\O(P_t)} = \frac{N^{-2}}{N^{-\frac{2}{3}}} = N^{-\frac{4}{3}}
\end{equation}

Therefore, the ratio scales as $\O(\Omega)= N^{-4/3}$.

\newpage
\subsection{Two-dimensional network}
In the case of two dimensions, such as the ones on the networks of the main paper, we need to approximate the scaling of the edges.
This is due to the optimized conductivities making the graph sparse, resulting in a linear scaling of the edges.

Similarly to \ref{subsec:one-dim-scaling}, we want to understand the scaling of the terms of the power by analyizing how its components scale with $N$.

Here, the scaling of the flow $\O(F^2)$ is unknown because it depends on the network structure and therefore $\gamma$.
Empirically $\O(F^2) \approx 1/N$ if $\gamma = 1/2$ and $\O(F^2) \approx 1/\sqrt{N}$ if $\gamma = 1/4$.
This is approximate because in reality $F^2$ does not scale uniformly: some edges have $\O(1)$ and some have $\O(1/N)$.

\vspace{2em}
Here we assume $\gamma = 1/2$ and take
\begin{align}
    \O(F^2) = N^{-1}
\end{align}
The remaining terms have known scaling with $N$:
\begin{align}
     &C = \left( \frac{F^2}{c_t \gamma} \right)^{\frac{1}{1 + \gamma}} \Rightarrow \O(C) = \O(c_t)^{-\frac{2}{3}} \, N^{-\frac{2}{3}}
     \\
     &\O(L) = N^{-\frac{1}{2}}
     \\
     &\O(\langle \ell \rangle) = N^{-\frac{1}{2}}
     \\
     &\O(\sum) = N
     \\
     &\O(s^2) = N^{-2}
\end{align}

So we find
\begin{equation}
    \O(P_t) =\O(\Sigma)\ \O(L)\ \O(F^2)\ \O(C^{-1})  +  \O(\Sigma)\ \O(L)\ \O(C^{1/2}) = N^{\frac{1}{6}} +  N^{\frac{1}{6}} = N^{\frac{1}{6}}
\end{equation}

and analogously
\begin{align}
\O(P_d) = \O(\Sigma)\ \O(\langle \ell \rangle)\ \O(s^2) =  N^{-3/2}.
\end{align}

Therefore, the resulting power ratio scales as 
\begin{equation}
    \O(\Omega) = \frac{\O(P_d)}{\O(P_t)} = \frac{N^{-3/2}}{N^{1/6}} = N^{-10/6} = N^{-5/3}
\end{equation}
which changes the scaling w.r.t. the one dimensional case.

\newpage

\section{Differentiable Voronoi}

In this section, we show an overview of the clipped Voronoi tessellation implementation used.
Notably, our approach to make it differentiable relies on a pre-calculation using \textsc{SciPy}, and leveraging that information to efficiency construct the resulting diagram.

\begin{algorithm}[H]
\caption{Differentiable Voronoi Tessellation Clipped by a Boundary}
\begin{algorithmic}[1]
\Require Set of points $x = \{(x_i, y_i)\}$, convex boundary polygon $\text{boundary}$
\Ensure Voronoi nodes and regions clipped within the boundary

\Procedure{differentiable\_voronoi}{$x$, $\text{boundary}$}

    \State \textbf{Voronoi Pre-calculation (non-differentiable using SciPy and CPU)}
    \State Extend $x$ by scaling the boundary to avoid infinite regions:
        \[
        x_{\text{ext}} = x \cup \left(\text{boundary} + 100 \cdot (\text{boundary} - \text{mean}(x))\right)
        \]
    \State Compute initial Voronoi regions and Delaunay triangulation using \texttt{scipy.spatial}.
    \State Identify regions needing clipping by checking intersections of Voronoi edges with boundary edges.

    \vspace{2em}
    \State \textbf{Differentiable Clipped Voronoi (GPU friendly)}
    \State Use Delaunay triangulation on $x_{\text{ext}}$ to find Voronoi nodes as follows:
    \For{each triangle defined by points $\{(a_x, a_y), (b_x, b_y), (c_x, c_y)\}$}
        \State Compute determinant:
            \[
            D = 2 \cdot ((a_x - c_x) \cdot (b_y - c_y) - (b_x - c_x) \cdot (a_y - c_y))
            \]
        \State Compute coordinates $(u_x, u_y)$ of the Voronoi node:
            \[
            u_x = \frac{(a_x^2 + a_y^2)(b_y - c_y) + (b_x^2 + b_y^2)(c_y - a_y) + (c_x^2 + c_y^2)(a_y - b_y)}{D}
            \]
            \[
            u_y = \frac{(a_x^2 + a_y^2)(c_x - b_x) + (b_x^2 + b_y^2)(a_x - c_x) + (c_x^2 + c_y^2)(b_x - a_x)}{D}
            \]
    \EndFor

    \State Determine which regions require clipping based on intersection and inclusion conditions.
    \For{each Voronoi region}
        \If{region needs clipping}
            \State Clip the region to the boundary by finding intersections of edges $(A, B)$ and $(C, D)$ as follows:
            \State Calculate intersection parameter:
                \[
                u_a = \frac{(D - C) \times (A - C)}{(B - A) \times (D - C)}
                \]
            \State If $0 \leq u_a \leq 1$, compute intersection point:
                \[
                (x, y) = A + u_a \cdot (B - A)
                \]
            \State Sort clipped vertices by angle $\theta$ relative to the centroid for ordered polygon representation:
                \[
                \theta = -\arctan2(y - y_{\text{centroid}}, x - x_{\text{centroid}})
                \]
        \EndIf
    \EndFor

    \State \textbf{Return Voronoi Nodes and Regions}
    \State Concatenate and reorder clipped regions to form the final set of valid Voronoi polygons within the boundary.
\EndProcedure
\end{algorithmic}
\end{algorithm}
Since this implementation is written with GPU in mind, most of the loops are instead vector operations, and calculation occur on its majority in parallel.
For calculating the gradients we rely on jax~\cite{jax2018github} automatic differentiation rules.

\newpage

\section{Implicit Function Theorem}

During the optimization of the nodes positions, we perform a full optimization of the conductivities of the system.
To do so, we use the dynamical model defined in the main text as

\begin{equation}
    \label{eq:ode-c}
    \dfrac{\mathrm{d} C_e}{\mathrm{d} t} = \left( \dfrac{F^2_e}{C_e^{\gamma+1}} - \gamma c_t \right) C_e + c_0 e^{-\lambda t},
\end{equation}
until it has reached the steady state $\boldsymbol{C}^\ast$.
The steady state solution is equivalent to finding the fixed point of the equation $\frac{\mathrm{d} C_e}{\mathrm{d} t}=0$.
From now on, we are interested in solving for the implicit function $f(C^\ast, x)=0$.

\vspace{1em}
The result $\C^\ast$ is dependent on the positions of the nodes $\x$ in a non-trivial manner.
Namely, $\C^\ast$ depends on $\boldsymbol{F}_e$, which is a function of $\boldsymbol{L}_e$ and $\boldsymbol{s}$, all of which dependent on $\x$.
We are interested in finding $\frac{dC}{dx}$, and even though automatic differentiation would allow to apply the chain rule through the iterative solver, the memory and computational cost of calculating derivatives and storing them would make it computationally challenging.
Luckily, since we are iterating over an implicit function, we can make use of the implicit function theorem~\cite{bell_algorithmic_2008, bertsekas_nonlinear_2016}, which allows for backpropagation through a fixed-point iteration without requiring saving each iterative step into memory.

For simplicity of presentation, here we will consider that the power only depends on the optimal conductivities, i.e.
\begin{equation}
    \frac{dP}{dx} = \frac{d P}{d C^\ast} \cdot \frac{d C^\ast}{d x}.
\end{equation}

In order to compute $\frac{dC^*}{dx}$ (the gradients of interest here), we use its implicit form $f(C, x)=0$ and obtain

\begin{equation}
    \frac{df}{dx} = \frac{\partial f}{\partial C} \cdot \frac{d C^*}{d x} + \frac{\partial f}{\partial x} = 0.
\end{equation}

Letting $A = \frac{\partial f}{\partial C}$ (the Jacobian of $\boldsymbol{f}$ with respect to $\boldsymbol{C}$)  and $B = \frac{\partial f}{\partial x}$ (the Jacobian of $\boldsymbol{f}$ with respect to $\boldsymbol{x}$), we isolate $\frac{\partial C^\ast}{\partial x}$:
\begin{equation}
    \frac{dC^\ast}{dx} = -A^{-1} B.
\end{equation}
This Jacobian is memory intensive (here we are ignoring the dependencies on the solver of $\boldsymbol{F}$).
However, we can rephrase this as a fixed-point problem using the adjoint vector $w$ given by

\begin{equation}
    w^T = v^T A.
\end{equation}
where $v = \frac{\partial f}{\partial C^\ast}$.
This means that we can solve for $w$ iteratively if we express it in implicit form such that $w = f(w) = w^T - v^tA$  where we are around $f(w) = 0$, and have
\begin{equation}
    \frac{df}{dx} = w^T B.
\end{equation}
which is equivalent to finding $\frac{dC}{dx}$ in the regime where $f(C, x) = f(C^\ast, x) = 0$.
Thus, by applying the implicit function theorem, we can compute the gradient with respect to parameters in the adaptation model for conductivity optimization, ensuring efficiency by avoiding the storage and backpropagation through each individual iteration step.

Our implementation of the fixed point approach can be found at~\url{https://github.com/kirkegaardlab/gradnodes}.
\end{document}